\documentstyle[12pt]{article}

\newcommand{\be}{\begin{equation}}
\newcommand{\ee}{\end{equation}}
\newcommand{\bq}{\begin{eqnarray}}
\newcommand{\eq}{\end{eqnarray}}
\newcommand{\nl}{\newline}

\begin{document}

\begin{flushright}
QMW-PH-96-9
\end{flushright}
\begin{center}

{\Large \bf $M$-theory Compactification \nl and Two-Brane/Five-Brane 
Duality.}\nl

{\Large B.S.Acharya\footnote{e-mail:r.acharya@qmw.ac.uk.
Work supported by PPARC.}}
{\it Dept. of Physics, Queen Mary and Westfield College, Mile End Road,
London, E1 4NS.}

\begin{abstract}
We discuss various dual {\it pairs} of
$M$-theory compactifications.
Each pair consists of a compactification in which the two-brane
plays the central role in relation to a string theory and a
compactification in which the five-brane takes center stage.
We show that in many examples such dual pairs
are interchanged by the {\it same} duality transformation in
each case.

\end{abstract}
\end{center}

\newpage
\section{Introduction.}

Orbifolds of $M$-theory have recieved a substantial amount of attention
in recent months [1-8]. These studies have proved crucial in
formulating links between the various string theories and $M$-theory itself.
This paper is devoted to studying further orbifolds of the eleven dimensional
theory, connecting these orbifolds to string theories and gaining a small
insight into the properties of $M$-theory itself. 

Specifically, we will be interested in the relationship between the two-brane
and five-brane of $M$-theory and the roles they play in compactifications
which are related to string theories in an appropriate limit.
Evidence is mounting that there exists a duality between different
compactifications of $M$-theory, which not only acts on the compactification
space, but exchanges two-branes with five-branes \cite{25}. 

One example
of this is $M$-theory on ${S^1}/{Z_2}{\times}T^3$ and $K3$. Both of
these are believed to be equivalent to the heterotic string on $T^3$
\cite{hor,w1}. 
In fact, the strong coupling
limit of the heterotic string in seven dimensions is described by {\it both}
of these $M$-theory compactifications (in the former case with a
long line segment, and with a $K3$ of large volume in the latter). In other
words, we have two eleven dimensional descriptions of the {\it same} physics.
It therefore seems possible that there 
exists a duality between
these compactifications of $M$-theory; after all, if our experience 
with string theory is anything to go by, two compactifications which
describe the same physics are usually related by a duality transformation.
We will refer to $M$-theory
compactifications which describe the ${\it same}$ string theories in an
appropriate limit, as dual compactifications. We will later give evidence
for a duality transformation which maps between such dual theories.

In $M$-theory on ${S^1}/{Z_2}{\times}T^3$ it is the two-brane wrapped around
the line segment which is identified with the fundamental heterotic string,
with a coupling constant proportional to a power of the segment length 
\cite{hor}. In the dual description ie on $K3$, the heterotic string may 
be identified with a five-brane wrapped around $K3$, with heterotic
coupling proportional to a power of the $K3$ volume \cite{w1,town}. 
If we make the
assumption that there is a duality transformation which interchanges these
two compactifications, then this tranformation appears to
exchange two-branes with five-branes. 

We will
be interested in giving further evidence for such a duality. We will give
some more examples of the situation in which one
finds {\it two} different $M$-theory compactifications which
are related to the {\it same} (or more generally, physically equivalent) 
string theory compactification.
In one of these pairs of $M$-theory compactifications
the two brane plays the central role, with the five-brane taking center stage
in the dual compactification. In fact, we will show that in the three dual
pairs of $M$-theory compactifications that we will discuss explicitly
(and also a much larger class which were discussed in \cite{Ach}, as
well as several of the models considered in \cite{vaf,stro}), the
transformation which maps between
the dual $M$-theory compactifications, 
is in fact the {\it same} transformation.
This makes the proposal for such a duality much more appealing.

The strategy for this will be to find two different compactifications of
$M$-theory which are related to the {\it same} string theory compactification.
As we will see, the particular string compactifications we discuss have some
interesting subtleties which have a very natural interpretation in {\it both}
$M$-theory compactifications, which is compelling evidence for the
proposed equivalences.

There are two properties of $M$-theory which will play a crucial role:\nl
{\it (i):} The two-brane and five-brane wrapped around $S^1$ and
$T^4$ respectively are both equivalent to an elementary Type IIA string (see
\cite{M} and refs. therein.).
Similarly, the two-brane and the five-brane wrapped around
${S^1}/{Z_2}$ and $K3$ respectively are both equivalent to an
elementary heterotic string \cite{hor,town}.\nl
{\it (ii):} In certain cases \cite{wit}, the ``twisted sector'' of orbifold
compactifications of the theory consists of two-branes \cite{sen1} 
and five-branes \cite{wit}.

Finally, because this work is closely related to recent work of Sen
\cite{sen1,sen2}, we will follow the notations introduced in \cite{sen1,sen2}.
These are the following:\nl
$I_n$ will denote the $Z_2$ isometry of an $n-torus$ which reflects all $n$
coordinates\footnote{When $n$ is odd we will specifically be
discussing $M$-theory or Type IIA compactifications. In these cases the
transformation $I_n$ must be combined with a transformation which
changes the sign of the three-form potentials in both these
theories. This will be understood in what follows.}. 
$F_L$ will denote the left moving fermion number operator in the
Type IIA or Type IIB string theories. $\Omega$ denotes the world-sheet parity
operator in the Type IIB theory. Finally, by abuse of notation, we will
denote orbifold isometry groups by their generators, eg instead of writing
${T^n}/{Z_2}$, we will write ${T^n}/{I_n}$.

In the next section, we will briefly review two orbifold compactifications
of $M$-theory which will play a central role in what follows. Then in
sections three and four we will give some examples of dual $M$-theory
compactifications to two dimensions. Some of the string theory analogues
of these compactifications were studied recently in \cite{sen2}. In
section five we show that many $M$-theory compactifications have a dual
description, with the role of the five-brane being interchanged with the
two-brane. In all the cases we have considered, these dual compactifications
are mapped into one another by the {\it same} duality transformation.

\section{$M$-theory on Two Orbifolds.}

In this section we will briefly review the relevant features of $M$-theory
compactified on ${T^n}/{I_n}$ for $n=5$ and $n=8$. These orbifolds were
discussed in \cite{wit,das,sen1} and this section is intended as a brief
review of some relevant features contained in those references.

{\it (i):} ${T^5}/{I_5}$

The massless bosonic fields of
$M$-theory are the metric and three-form potential. On ${T^5}/{I_5}$, the
untwisted sector massless spectrum ie the massless states of $M$-theory
on $T^5$ which survive the $I_5$ projection, consist of the six dimensional
chiral $N=2$ supergravity multiplet plus five tensor multiplets (which
contain anti self-dual two-forms). Anomaly cancellation requires a further
sixteen tensor multiplets from the ``twisted sector'' of the theory. It was
realised by Witten \cite{wit} that these multiplets arise naturally
as the world volume fields
of sixteen $M$-theory fivebranes located on the internal space.
Compelling evidence was given in \cite{wit}, 
that this compactification of $M$-theory, for a particular configuration
of the five-branes,
is equivalent to Type IIB string theory on $K3$, with a coupling constant
proportional to the radius of {\it any} of the circles of ${T^5}/{I_5}$. 
In particular, this leads to 
the conjecture that $M$-theory on ${T^5}/{I_5}{\times}X$ is equivalent, on
shrinking of the first factor, to Type IIB string theory on $K3{\times}X$,
with $X$ being any space. In this paper we will be interested in the cases
in which $X$ is $T^4$ or $K3$.

{\it (ii)} $M$-theory on ${T^8}/{I_8}$.

This $M$-theory compactification was discussed in \cite{sen1}. It was realised
in \cite{sen1} that the transformation $I_8$ gets mapped to a particular
transformation in the Type IIA theory on $T^7$. This is
${(-1)^{F_L}}.{I_7}$. However, this tranformation of the IIA theory
on $T^7$ is equivalent to the world-sheet parity transformation of the
Type IIB theory and the map between these two theories involves inverting 
the radii of all seven circles. 
Thus, it was conjectured in \cite{sen1} that $M$-theory
on ${T^8}/{I_8}$ is equivalent to Type IIB theory on ${T^7}/{\Omega}$. This is
by definition the Type I theory on $T^7$, and hence is equivalent by
Type I- heterotic duality \cite{hetI} to the heterotic string theory on $T^7$.
When viewed as an orbifold of Type IIB theory, the Type I 
theory has a ``twisted sector''
spectrum consisting of $32$ nine-branes \cite{pol}. 
Now, retracing these nine-branes
back to the Type IIA theory involves inverting the radii
of all seven circles of $T^7$
thus converting these nine-branes into $32$ two-branes. However, from
the Type IIA point of view, of these $32$ two-branes, only $16$ are
independent, due to the action of the orbifold group. Finally, since
a two-brane of the Type IIA theory is also a two-brane of $M$-theory, we
see that the ``twisted sector'' spectrum of $M$-theory on ${T^8}/{I_8}$
is sixteen two-branes moving on the internal orbifold. Each of these
two-branes carries one vector multiplet of the $N=8$ supersymmetry of
the three dimensionsal theory \cite{sen1}.

We will later be interested in a further reduction of this $M$-theory 
compactification on $S^1$ and ${S^1}/{I_1}$.

\section{Examples of Dual $M$-theory Compactifications.}

{\it (i)}: $M$-theory on: $(A)=$ ${T^5}/{I_5}{\times}T^4$ 
and $(B)=$ ${T^8}/{I_8}{\times}S^1$

Consider $M$-theory compactified on both of these spaces. They are, 
respectively, toroidal compactifications to two dimensions of the two cases
considered in the previous section. Consider case $(A)$ first. Because we
are discussing compactifications to two dimensions, massless scalars 
decompose into left and right movers. We will denote by $(L,R)$ a two
dimensional compactification with $L$ left moving scalars and $R$ right
moving ones. 

In case $(A)$, the untwisted sector spectrum of $M$-theory is just
the spectrum of $M$-theory on $T^9$ which is invariant under $I_5$.
This is just the $N=(8,8)$ supergravity multiplet with a scalar
field content of $(64,64)$ which together with their
superpartners form representations of the 
supersymmetry. The ``twisted sector'' of this theory is just the
twisted sector of $M$-theory on ${T^5}/{I_5}$, toroidally compactified
to two dimensions. As realised by Witten \cite{wit}, and as reviewed
above, the ``twisted sector'' consists of $16$ five-branes. Now, a five-brane
of $M$-theory wrapped on $T^4$ is just an elementary Type IIA string. 
Each of these strings contributes $(8,8)$ to our spectrum of massless
fields, giving a total of $(192,192)$ scalars.

According to \cite{wit} the $M$-theory compactification we have just
discussed should go over to Type IIB on $K3{\times}{T^4}$, upon shrinking of
one of five circles. It is straightforward to check that this compactification
of Type IIB theory gives the correct spectrum. Further, this
compactification of Type IIB is equivalent by T-duality to the
Type IIA theory on ${K3}{\times}{T^4}^{\prime}$, which is in turn equivalent
\cite{w1} to the heterotic string on an eight torus, $T^8$.

Very recently, it was realised by Sen \cite{sen2} that the Type IIB theory
on $K3{\times}T^4$ is dual, by a sequence of duality transformations, to
Type IIA string theory on ${T^8}/{I_8}$. As discussed in \cite{loop} 
certain compactifications of Type IIA and heterotic
strings to two dimensions contain some subtleties due to the existence of
one-loop tadpoles associated with the $NS$ sector two-forms. It was
argued in \cite{sen2} and commented upon in \cite{F}
that the inconsistencies associated with such
tadpoles can be cured by the introduction of $n$ elementary strings, where
$n$ is an integer which characterises the ``extent'' of the tadpole
inconsistency. Now, as we discussed above, from the $M$-theory point
of view, we find $(64,64)$ states in the untwisted sector together
with $16{\times}(8,8)$ which are carried by elementary Type IIA strings which
come from the $16$ five-branes wrapped around $T^4$. We therefore
can expect to see these elementary strings present in the Type IIA
theory on ${T^8}/{I_8}$. In fact this orbifold of Type IIA
theory was analysed by Sen \cite{sen2}, where he showed that
precisely $16$ elementary IIA strings are required to cure the
one-loop tadpole inconsistency. In the untwisted sector, we find the
$(64,64)$ states which were present in the untwisted sector of the $M$-theory
compactification. Thus, this Type IIA compactification also has
the same spectrum and it is indeed compelling that the extra strings
are required not only from the $M$-theory point of view, but for consistency
of the Type IIA theory itself.

Now let us consider case $(B)$ ie $M$-theory on ${T^8}/{I_8}{\times}S^1$.
We reviewed the case of $M$-theory on ${T^8}/{I_8}$ in the
previous section. Thus all that remains is to reduce this model on $S^1$.
In the untwisted sector, we again find the $(64,64)$ spectrum of
massless states. The twisted sector consists of sixteen two-branes wrapped
on $S^1$, which are equivalent to sixteen Type IIA strings. However, this
compactification of $M$-theory is equivalent to Type IIA theory on
${T^8}/{I_8}$, which as we saw above, also has the same spectrum.

We have thus given evidence that $M$-theory on ${T^5}/{I_5}{\times}T^4$
is physically
equivalent to Type IIB on $K3{\times}T^4$, the heterotic/Type I theory
on $T^8$, and finally the Type IIA theory on ${T^8}/{I_8}$. In this
$M$-theory compactification, the five-brane played a crucial role.
Secondly, we have seen that in $M$-theory on ${T^8}/{I_8}{\times}S^1$,
two-branes play a crucial role. Further, this theory is {\it also}
equivalent to Type IIA on ${T^8}/{I_8}$. This strongly suggests that there is
a symmetry of $M$-theory in which five-branes are interchanged with
two-branes. We will identify such a symmetry after the next section.

We now go on to discuss in a similar fashion two more
$M$-theory compactifications, both of which appear to be
equivalent to a {\it single} compactification of heterotic string
theory. Again, we will see that in one case the five-brane plays
a crucial role, whereas in the other case it is the two-brane which does
so. 

\section{Further Examples.}

In this section we will again consider $M$-theory compactified on
two different spaces. These examples will be very similar to
the cases considered in the previous section, except we will replace
the $T^4$ factor in case $(A)$ with $K3$ and the $S^1$ factor in
case $(B)$ with ${S^1}/{I_1}$. We therefore expect to see the elementary
Type IIA strings being replaced with elementary heterotic strings.

{\it (i):} $M$-theory on ${T^5}/{I_5}{\times}K3$.

This 
compactification should be equivalent to the Type IIB string theory on
$K3{\times}K3$ and also to a particular orbifold of the heterotic
string in two dimensions \cite{wit}.

In the $M$-theory case, we must simply reduce the model of \cite{wit} on
$K3$. Let us first analyze this model as an orbifold of $M$-theory on
${T^5}{\times}K3$. In the untwisted sector, we find the massless states
in the effective two-dimensional theory comprise the chiral $N=(8,0)$
supergravity multiplet and (192,192) scalars. The twisted sector of
this model consists of the twisted sector of the model considered in 
\cite{wit}, reduced on $K3$. As noted above, the twisted sector of the model
in \cite{wit} consisted of sixteen fivebranes. The bosonic world volume field
content of the fivebrane consists of an anti self-dual two-form potential
and five scalars. Thus, to calculate the spectrum of our model, we
simply need to wrap these sixteen
fivebranes on $K3$. A fivebrane
in $M$-theory wrapped on $K3$ is equivalent to a string with the world sheet 
field content of the heterotic string \cite{town}.
Thus in the
notation above, the double dimensional reduction of the fivebrane
on $K3$ gives rise to (24,8) scalars in the two dimensional theory. These are
of course the world sheet scalar degrees of freedom of the heterotic string
in light cone gauge. Thus, as we have 16 fivebranes in six dimensions, 
the massless scalar content of this $M$-theory
compactification consists of (192,192) + 16(24,8) = (576,320). Given the
number of supersymmetries, it can be checked that the full spectrum is
free from anomalies. We will shortly see how these sixteen elementary
heterotic strings naturally arise in an orbifold of the heterotic
string compactified to two dimensions.

According to \cite{wit}, the above $M$-theory compactification should
be equivalent to the Type IIB string on $K3{\times}K3$. In ten dimensions,
the massless bosonic field content of the Type IIB theory consists of
two scalars, two 2-form potentials, a 4-form potential and the metric. The
4-form potential is self-dual. The product manifold $K3{\times}K3$ has the
following non-zero Betti numbers: ${b_0}=1$, ${b_2}=44$, 
${b_4^+}=371$, ${b_4^-}=115$, ${b_6}=44$ and ${b_8}=1$. Reducing the two 
2-forms on this manifold gives rise to (88,88) scalars. Reducing the
self-dual four form gives (371,115) scalars. Because it is possible
to define a torsion free $Spin(7)$ structure on {\bf $R^8$} which is
preserved by the holonomy structure of $K3{\times}K3$,
the moduli space of metrics on
$K3{\times}K3$ has dimension ${b_4^-} + 1$$=116$ \cite{J3}. Equivalently, each
$K3$ has a moduli space of metrics of dimension $58$, giving $116$ for the
product.
However, one of these components
becomes part of the supergravity multiplet, as the metric has formally -1
degrees of freedom in two dimensions. This means that the metric contributes
(115,115) scalars to the 2d theory. Finally, each ten dimensional scalar
decomposes into a left moving scalar and a right moving one. All
in all, the field content of this theory is precisely that of the
$M$-theory compactification.

A more difficult problem is to relate this $M$-theory compactification
to a compactification of heterotic string theory. If we exchange the two 
factors in the above $M$-theory compactification, we have $M$-theory on
$K3{\times}{T^5}/{I_5}$. Now, $M$-theory on $K3{\times}{T^5}$ should \cite{w1}
be equivalent to the heterotic string on ${{T^3}{\times}T^5}$. Thus, in
the compactification we are discussing one would naively expect that the
heterotic compactification is on ${T^3}{\times}{T^5}/{I_5}$. However, as
the heterotic string is an oriented string theory, it is not possible to
formulate the theory on this background\footnote{These arguments were made
in a similar context in \cite{duff}.}. We thus expect that the heterotic
dual compactification should be some other $Z_2$ orbifold of the heterotic
string on $T^8$. 

According to \cite{sen1} the transformation $I_5$ in $M$-theory
gets mapped to ${(-1)^{F_L}.I_4}$ in the Type IIA theory on $T^8$ \cite{sen1}
\footnote{The following derivation of the heterotic dual of the Type IIA
model was also given in \cite{sen2}.}.
We thus expect that the $M$-theory compactification
we are discussing is equivalent to the Type IIA theory on a $Z_2$ orbifold
of $K3{\times}T^4$, where the $Z_2$ element is the one just described.
It is possible to check that the untwisted and twisted sector spectra are the
same in the $M$-theory and IIA theories. This compactification of
Type IIA theory can now be mapped to a heterotic compactification.

On $K3$, the Type IIA theory is believed to be equivalent to the heterotic,
string on $T^4$, which has Narain lattice, ${\Gamma}^{20,4}$. 
The transformation, ${(-1)^{F_L}}$ maps to the $Z_2$ transformation
which inverts ${\Gamma}^{20,4}$ in the heterotic description \cite{vaf}.
The Type IIA compactification which gives the same spectrum as the
$M$-theory background which interests us is mapped in the heterotic
theory to a $Z_2$ orbifold of ${\Gamma}^{24,8}$, which describes the
heterotic string in two dimensions. The transformation ${{(-1)}^{F_L}}$ in
the Type IIA theory then corresponds to inversion of a signature $(20,4)$
factor in this lattice. The transformation which inverts all the coordinates
of the $T^4$ factor in the Type IIA compactification is mapped to a reflection
of the remaining signature $(4,4)$ piece of the lattice, since this $T^4$ is
``common'' to both the Type IIA and heterotic compactifications. This
leads us to the final statement that the $Z_2$ transformation which
defined our original orbifold in $M$-theory gets mapped to the following
transformation in the heterotic string, toroidally compactified to two
dimensions:

\be
g: {\Gamma}^{24,8} = -{\Gamma}^{24,8}
\ee

ie this transformation inverts the entire lattice of the compactified string
theory.
We should therefore, perhaps naively, expect that our $M$-theory 
compactification is equivalent to a $Z_2$ orbifold of this heterotic 
compactification with orbifold group generated by $g$. 
In particular, this orbifold must satisfy several criteria:
(i) it must be modular invariant.
(ii) it should give the correct supersymmetry algebra in the effective
two dimensional theory, namely $N=(8,0)$.
(iii) it should give the same field content as the above theories.

It is straightforward to check that points (i) and (ii) are satisfied. To
verify point (iii) additional subtleties are involved. 
Firstly, in the untwisted
sector, the massless spectrum of the heterotic orbifold is precisely that
which we found above in the $M$-theory case. One would therefore expect that
the twisted sector should contain the requisite states that we found in
$M$-theory. However, because the left moving twisted sector ground state
is massive, we will not find any massless states in the twisted sector - at
least none which arise in the standard fashion as tensor products of left
and right movers on the twisted worldsheet. However, the untwisted spectrum
that we have in this model is the {\it same} as that of $M$-theory on
$K3{\times}{T^5}/{I_5}$, namely $(192,192)$ scalars. Further, because we
saw that in the $M$-theory case, the twisted sector consisted of $16$
elementary heterotic strings, it is natural to expect
that they are also present
here. In the heterotic compactification we are discussing, it is 
therefore again natural
to expect that the presence of these strings cures any inconsistencies in
the one-loop tadpoles \cite{loop}
that may be present. In fact, this particular
orbifold of the heterotic string was also analysed by Sen \cite{sen2}, and
it was found that precisely sixteen elementary heterotic strings are
required to cancel any inconsistencies. This is of course compelling
evidence that the theories are equivalent. We will now consider
another $M$-theory compactification which is equivalent to the
heterotic orbifold which we have just discussed.

{\it (ii):} $M$-theory on ${T^8}/{g^{\prime}}{\times}{S^1}/{I_1}$.

Our aim is to construct an orbifold of $M$-theory which is equivalent to
the heterotic orbifold we have just considered. Because the $E_8{\times}E_8$
heterotic string in ten dimensions is believed to be equivalent to
$M$-theory on ${S^1}/{I_1}$ we expect that symmetries of the heterotic
string be present in this $M$-theory compactification. Thus, since
our goal is to find another $M$-theory compactification which is
equivalent to the heterotic orbifold of the previous section, it should
be possible to {\it define} the required orbifold of $M$-theory 
{\it a priori}. This is precisely the strategy we will take.

Since the heterotic orbifold which interests us is an orbifold of the
toroidally compactified heterotic string in two-dimensions, we must
first consider $M$-theory on ${S^1}/{I_1}$ toroidally compactified to
two dimensions. Such a compactification of $M$-theory contains two ten
dimensional $E_8$ Super-Yang-Mills multiplets toroidally
compactified to two dimensions. The full scalar field content of the
compactification is $(192,192)$. Of these, $(128,128)$ originate
from the ten-dimensional vector multiplets. In the heterotic orbifold which
we wish to ``duplicate'' in $M$-theory, the action on $T^8$ is simply
$I_8$. Thus, we expect $g^{\prime}$ to contain such a factor. Further, as
we reviewed above, the ``twisted sector'' associated with $I_8$ is
sixteen two-branes, which when reduced on ${S^1}/{I_1}$ are precisely the
sixteen heterotic strings which were required for consistency of the 
heterotic compactification. It therefore becomes plausible that $g^{\prime}$,
which defines the $M$-theoretic analogue of $g$, is simply $I_8$.
On ${T^8}/{I_8}{\times}{S^1}/{I_1}$, $M$-theory
contains $(64,64)$ massless scalars which originate from the metric and
three-form. However there exists the possibility that the modes associated
with the ten-dimensional vector multiplets will contribute. As we
mentioned above, 
before the $I_8$ projection there are precisely $(128,128)$ such
states. By symmetry, because $I_8$ acts identically on all $1-cycles$ of $T^8$,
we expect that either these $(128,128)$ modes are odd under $I_8$ or
they are all even. In the case when these states are odd, one can check that
the corresponding heterotic string orbifold does {\it not} give a spectrum
which matches the $M$-theory one. In the case when the $(128,128)$ states are 
even, $I_8$ has precisely the same action in $M$-theory as $g$ has on the
heterotic string and we find complete agreement between the spectrum of states
in the two compactifications. To recap, this is namely
$(192,192)$ from the untwisted sector plus $16\times(24,8)$ which are carried
by elementary heterotic strings. From the $M$-theory point of view these
strings
come from two-branes wrapped around the line segment. Thus, we expect
that the $M$-theoretic
analogue of $g$ is $I_8$.

We have thus related the heterotic string on the orbifold defined
by $g$ to $M$-theory on $K3{\times}{T^5}/{I_5}$ and $M$-theory on
${T^8}/{I_8}{\times}{S^1}/{I_1}$. 

In the
first of these cases, we saw that 
the elementary heterotic string may be identified as
a five-brane of $M$-theory wrapped around $K3$ and the $16$ additional
heterotic strings required for consistency of the string theory have
an $M$-theoretic interpretation as the $16$ five-branes of \cite{wit}
wrapped on $K3$. 

In the second of these cases the elementary heterotic string is naturally
identified with a two-brane of $M$-theory wrapped around ${S^1}/{I_1}$ with
the additional $16$ requisite heterotic strings being interpreted
as the $16$ two-branes of \cite{sen1} wrapped around ${S^1}/{I_1}$.

\section{Duality Symmetry.}

In this concluding section we will identify
the $Z_2$ transformation which defines a duality map between an
$M$-theory compactification in which the two-brane plays the crucial role in
the relation to a string theory and a compactification in which the five
brane does so. The strategy we will follow will be to identify the required
transformation in the Type IIA theory which will then allow us to induce
the requisite transformation in $M$-theory. We will find that the {\it same}
transformation defines the map between all the pairs of $M$-theory
compactifications which we have discussed in this paper as well as 
a much wider class which were discussed in \cite{Ach} and also several of 
the models considered in \cite{vaf,stro}. We take this as 
further evidence that $M$-theory possesses such a duality symmetry.

In \cite{sen1}, strong evidence was presented that for $M$-theory on
$X{\times}{S^1}$, with $X$ any space, 
the transformation $I_1$, {\it if} 
combined with a transformation $\alpha$ which 
acts on $X$, is represented
as $(-1)^{F_L}.{\alpha}$ in the Type IIA theory on $X$. For the case when
$X$ is a torus, $T^n$, we can identify the transformation
$I_1$ in $M$-theory as the transformation $I_1$ in the Type IIA
theory on ${T^n}^{\prime}$ \cite{hor}. We will assume that these properties 
hold
in general in $d<10$\footnote{In $d=10$ it was shown in \cite{sen1} that this
does not hold.}. 
Thus if we consider $M$-theory on $T^n$, we
can rewrite the transformation $I_n$ as ${I_1}{\times}I_{n-1}$. This then
translates to the Type IIA theory on $T^{n-1}$ as the transformation
$(-1)^{F_L}{\times}I_{n-1}$. Thus, in certain cases, we
may identify the transformation $(-1)^{F_L}$ in the Type IIA
theory with $I_1$ in $M$-theory. Another fact that we will require is that
the Type IIA theory on $T^4$ has a self duality transformation $\sigma$ which
takes $(-1)^{F_L}$ into $I_4$ \cite{sen2}. These observations will
allow us to identify the transformation which maps between the pairs of
$M$-theory vacua that we have discussed above.

{\it (i)}: $M$-theory on: $(A)$: ${T^5}/{I_5}{\times}T^4$ 
and $(B)$: ${T^8}/{I_8}{\times}S^1$

In these examples, we identified a dual Type IIA comapactification on
${T^8}/{I_8}$. However, upon rewriting $T^8$ as ${T^4}^{\prime}{\times}T^4$ 
and
$I_8$ as ${I_4}^{\prime}I_4$ and
using the self duality transformation $\sigma$, we can map this Type IIA
theory to the Type IIA theory on 
$({{T^4}^{\prime}\times{T^4}})/{{I_4}^{\prime}.{(-1)^{F_L}}}$.
In this example, because $(-1)^{F_L}$ is combined with another 
transformation, we can identify its $M$-theory analogue as $I_1$. Now, in
case $(A)$ above, we can see that the transformation $I_5$ $={I_1}.{I_4}$
is represented in the Type IIA theory on ${T^4}^{\prime}{\times}T^4$ as
$(-1)^{F_L}.I_4$. $\sigma$ in the Type IIA theory maps $(-1)^{F_L}$ to
${I_4}^{\prime}$. Hence,  $(-1)^{F_L}.I_4$ is mapped into 
${I_4}^{\prime}.I_4$$=I_8$. Finally, $I_8$ in the Type IIA theory
is just $I_8$ in $M$-theory, which is just case $(B)$ above. Thus if we
translate the transformation $\sigma$ into its $M$-theory counterpart, we
have a transformation which maps $I_1$ in case $(A)$ to a transformation
${I_4}^{\prime}$. In other words, the transformation $I_5$ in $M$-theory
on $T^9$ is mapped to the transformation $I_8$, by the $M$-theoretic
analogue of $\sigma$ in the Type IIA theory. We will denote this
transformation in $M$-theory as ${\sigma}_M$ for definiteness.

Using this information, we can now check if indeed ${\sigma}_M$ has a
similar action on the other pairs of examples that we considered. These
were:\nl
{\it (ii)}: $M$-theory on $(C)$: ${T^5}/{I_5}{\times}K3$ and
$(D)$: ${T^8}/{I_8}{\times}{S^1}/{I_1}$.

It is natural to restrict ourselves to the case when $K3$ is the orbifold
${T^4}/{I_4}$, otherwise it is not presently possible to perform
the required analysis.

In case $(C)$, the orbifold isometry group which acts on $M$-theory on $T^9$
has non-identity elements: $I_5$ and $I_4$. Under the action of
${\sigma}_M$ these transform as follows: 
\be
{{\sigma}_M}: {I_5} \rightarrow {I_8}
\ee
\be
{{\sigma}_M}: {I_4} \rightarrow {I_1}
\ee
Hence, the generators in case $(C)$ are transformed to those in case $(D)$.
Thus, ${\sigma}_M$ also defines a duality map between these two physically
equivalent compactifications of $M$-theory.

{\it (iii)}: $M$-theory on: $(E)$: ${T^5}/{I_4}$ and $(F)$: ${T^5}/{I_1}$.
These two cases are $S^1$ compactifications of the dual pair we discussed
in the introduction. Case $(E)$ is a special case of $M$-theory on
$K3{\times}S^1$, which we expect to be equivalent to the heterotic string
on $T^4$. Case $(F)$ is just $M$-theory on ${S^1}/{I_1}{\times}T^4$, which we
also expect \cite{hor} to be equivalent to the heterotic string on $T^4$.
In case $(E)$, the transformation $I_4$ can be identified with
the transformation $(-1)^{F_L}.I_3$ in the Type IIA theory on $T^4$ 
\cite{sen1}. By radius inversion on the three circles on which $I_3$ acts
this theory is mapped to Type IIB theory on ${T^4}/{\Omega}$, which is
just Type I theory on $T^4$, which is equivalent \cite{hetI} to the
heterotic string on $T^4$. $\sigma$ in the Type IIA theory on $T^4$ maps
$(-1)^{F_L}.I_3$ to ${I_4}{I_3}$$=I_1$. From the $M$-theory point of
view, $I_1$ in the Type IIA theory on $T^4$ is also $I_1$ in $M$-theory 
\cite{hor}.
Therefore, case $(E)$ is mapped to case $(F)$, by the action of
${\sigma}_M$.

{\it {iv}}: A Larger Class of Examples.

In \cite{Ach} we presented evidence that $M$-theory compactifications
on Joyce manifolds of $G_2$ and $Spin(7)$ holonomy are dual to 
compactifications of heterotic string theory on Calabi-Yau manifolds
and Joyce manifolds of $G_2$ holonomy respectively. These theories
are $N=1$ theories in four and three dimensions respectively. In these
cases, the five-brane of $M$-theory wrapped around a $K3$ submanifold
of the Joyce manifold may be identified with the fundamental heterotic
string. However, we can also expect that
there exists a dual $M$-theory compactification in which the wrapped
two-brane plays the fundamental role, since $M$-theory on
${S^1}/{I_1}{\times}X$ is equivalent to the heterotic string on $X$ 
\cite{hor}. 

In fact, all the compact manifolds discussed in \cite{Ach} were constructed
by Joyce as blown up toroidal orbifolds. Further, the only non-freely
acting orbifold generators in $M$-theory, were of the form $I_4$. Thus,
by applying the transformation ${\sigma}_M$ once, one of the generators
$I_4$ is mapped to $I_1$, as we discussed in the previous case. Although
we do not present the details here, it may be checked that in all
the $M$-theory compactifications in \cite{Ach} (for which the corresponding
heterotic dual is a compactification on some manifold $X$), the transformation
${\sigma}_M$ maps the $M$-theory compactification on the Joyce manifold
to $M$-theory on ${S^1}/{I_1}{\times}X$. Thus, in all the 
cases considered in \cite{Ach} ${\sigma}_M$ maps
one $M$-theory compactification in which the five-brane plays the crucial
role, to one in which the two-brane does so. This reasoning also applies
to the cases considered in \cite{stro} and also to
several of the models considered in \cite{vaf}.

{\it (v)}: $K3$ Fibrations.

Compactifications of $M$-theory and string theories on manifolds which
admit $K3$ fibrations have played an important role in our understanding
of string theory dualities \cite{vaf,K3}. In such $M$-theory compactifications
it is the five-brane wrapped around the $K3$ fiber which one identifies
as the heterotic string. If we consider a point in the moduli space in which
the fiber is ${T^4}/{I_4}$ (assuming that such a point exists), then we can
apply the transformation ${\sigma}_M$ which exchanges the $K3$ fibers with
${S^1}/{I_1}{\times}T^3$ fibers. In other words, we
can apply the duality transformation ${\sigma}_M$ fibrewise in the
adiabatic limit \cite{vaf}. In the dual compactification the
two-brane wrapped around the line segment is identified with the heterotic
string. If this argument does apply in this case, then it appears that
compactifications of $M$-theory on manifolds which admit $K3$ fibrations 
also have a dual
description.

\section{Comments.}

We have seen that, in a large number of cases,
${\sigma}_M$ exchanges a given 
$M$-theory compactification with a dual compactification. 
Five-branes in one compactification
are replaced with two-branes in the dual compactification, and vice-versa.
It therefore appears that such a duality
symmetry is a fairly general property of $M$-theory compactifications.
However, because this duality is a property of $M$-theory 
{\it compactification},
it is not clear if such a duality has an eleven dimensional origin, although
this does remain an open possibility. A two-brane/five-brane duality in
eleven dimensions was suggested in the first reference of \cite{25}.

We hope that these results will be a small clue towards the formulation
of $M$-theory.

{\bf Acknowledgements.}
The author is extremely indebted to Chris Hull for guidance and 
encouragement, and to PPARC,
by whom this work is supported.


\begin{thebibliography}{12}
\bibitem{hor} P.Horava and E.Witten, Nucl.Phys. B460 (1996) 506.
\bibitem{wit} E.Witten, Nucl.Phys. B463 (1996) 383.
\bibitem{das} K.Dasgupta and S.Mukhi, hepth/9512196.
\bibitem{sen} A.Sen, hepth/9602010.
\bibitem{Ach} B.S.Acharya, hepth/9603033 and hepth/9604133.
\bibitem{sen1} A.Sen, hepth/9603113.
\bibitem{sen2} A.Sen, hepth/9604070.
\bibitem{kou} A.Kumar and K.Ray, hepth/9604164.
\bibitem{25} C.Hull and P.Townsend, Nucl.Phys. B438 (1995), 109;
R.Dijkgraaf, E.Verlinde, H.Verlinde, hepth/9603126, hepth
/9604055; O.Aharony, hepth/9604103 and refs. therein.
\bibitem{w1} E.Witten, Nucl.Phys. B443 (1995) 85.
\bibitem{M} J.Schwarz, Phys.Lett. B367 (1996) 97.
\bibitem{town} P.Townsend, Phys.Lett. B354 (1995) 247.
\bibitem{hetI} C.M.Hull, Phys.Lett. B357 (1995) 545. A.Dabholkar,
Phys.Lett. B357 (1995) 307. J.Polchinski and E.Witten, hepth/9510169.
\bibitem{pol} J.Polchinski, S.Choudhuri and C.Johnson, hepth/9602052 and
refs. therein.
\bibitem{loop} C.Vafa and E.Witten, Nucl.Phys. B447 (1995) 261.
\bibitem{J3} D.D.Joyce, Inv. Math. Vol.123 Fasc.3 (1996) 507.
\bibitem{vaf} C.Vafa and E.Witten, hepth/9507050.
\bibitem{stro} J.Harvey, D.Lowe and A.Strominger, Phys. Lett. B362 (1995) 65.
\bibitem{duff} M.Duff, R.Minasian, E.Witten, hepth/9601036.
\bibitem{F} C.Vafa and E.Witten, unpublished, as quoted in ref.21
\bibitem{F1} C.Vafa, hepth/9602022.
\bibitem{K3} A.Klemm, W.Lerche and P.Mayr, Phys. Lett. B357 (1995) 69.
\end{thebibliography}
\end{document}